\documentclass[]{article}

\usepackage[]{graphics}
\usepackage[]{amsmath}
\usepackage[]{amssymb}
\begin{document}
\title{A new bond fluctuation method for a polymer 
undergoing gel electrophoresis}
\author{Ryuzo Azuma ~and~ Hajime Takayama\\
\it Institute for Solid State Physics, The university of Tokyo\\
7-22-1 Roppongi, Minato-ku, Tokyo 106-8666, Japan}
\date{\today}
\maketitle
\begin{abstract}
We present a new computational methodology for the investigation of 
gel electrophoresis of polyelectrolytes. 
We have developed the method initially to incorporate sliding motion 
of tight parts of a polymer pulled by an electric field into the bond 
fluctuation method (BFM). Such motion due to tensile force over 
distances much larger than the persistent length is realized by 
non-local movement of a slack monomer at an either end of the tight 
part. The latter movement is introduced stochastically. 
This new BFM overcomes the well-known difficulty in the conventional 
BFM that polymers are trapped by gel fibers in relatively large 
fields. At the same time it also reproduces properly equilibrium 
properties of a polymer in a vanishing filed limit. The new BFM 
thus turns out an efficient computational method to study gel 
electrophoresis in a wide range of the electric field strength.
\end{abstract}


\section{Introduction}

Gel electrophoresis is a method which separates polyelectrolytes
such as DNA according to their length. 
This technique is an application of the phenomenon which polyelectrolytes 
exhibit different migration 
velocities in gel when they are pulled by an 
external electric field. 
Although in recent years many sophisticated methods such as 
pulsed-field techniques have been developed \cite{norden,zimmlev}, 
its underlying physics, even merely on steady field techniques, are 
not thoroughly understood. 

The concept of reptation, which was proposed by de Gennes
\cite{degennes}, has been incorporated into theoretical analysis of
gel electrophoresis. For example, the biased reptation model has 
succeeded to explain an empirical law of steady field experiments 
in the small field limit \cite{ldz,slanool1}, i.e., DNA mobility is 
proportional to reciprocal of its length \cite{nancy2}.
However such predictions by the reptation theory are only on 
properties associated with dynamics of averaged conformations 
of DNA.

It is now well known that DNA dynamics in electrophoresis is more 
complicated and interesting. The molecular-dynamics (MD) simulation 
by Deutsch \cite{deutsch} on a freely jointed chain in a 
two-dimensional space 
with obstacles substituted for gel fibers first demonstrated that 
the chain migrates through obstacles taking extended and collapsed 
conformations alternatively. Since this oscillatory behavior appears
as a steady state \cite{deuma}, the averaged conformation which the
reptation model predicts is not stable. 
Also experimentally, since fluorescent microscopy has been 
invented and improved, evidences showing this instability have 
increased as well, and now details of DNA motion itself are of 
current interest in the study of gel electrophoresis
\cite{smith,SchKov,Gurrieri,howard}. Among them  
the inch-worm like motion may be the most typical example which is
observed for large DNA \cite{oana}. 

In studying polymer dynamics,  numerical simulations have 
played important roles. One of them is the MD 
method, 
whose example was already mentioned above. Another powerful 
simulation is the bond fluctuation method (BFM), which is a Monte 
Carlo method utilizing description of a polymer(s) on a 
lattice \cite{carkre}.
Its efficiency in examining gel electrophoresis in a small field 
was already 
demonstrated, but at the same time, 
its difficulty when applied to the case in a large field was 
pointed out \cite{kremer}. 
The difficulty is that it takes huge MC steps (mcs's) for polymers 
once hanging on obstacles to get rid of them. 
As a result mobility in a large field becomes significantly smaller 
than in a relatively small field. This is an artifact of the 
numerical method since in actual experiments the mobility is observed 
to increase monotonically with increasing field \cite{hebe}. 
A reason of this difficulty 
is considered that tensile force between monomers, which are of 
fundamental importance when each part of the chain is
in nearly straight conformations, 
is not taken into account in the conventional BFM (c-BFM).

The purpose of the present paper is to report a new BFM (n-BFM) which we 
have developed to overcome the above-mentioned difficulty by 
introducing to the c-BFM new stochastic processes which 
simulate sliding motions caused by tensile force. 
The new method turns out to be able to reproduce qualitative aspects 
of gel electrophoresis phenomena in a wide range of the 
field. In large fields polymer motion is free from trapping by
obstacles and mobility monotonically increases (and tends to saturate)
with increasing field as we have expected. In small fields, on the
other hand, the results simulated by the new method coincide with
those simulated by the c-BFM if the time unit is properly scaled. This 
means that, in this field range, the stochastic process newly
introduced here contributes to the entropy effect due to polymer 
conformations similarly to what the c-BFM process does. 
Furthermore, in a certain limited range of 
field we have observed extended and contracted conformations of a 
polymer very frequently, though not quasi-periodically. 
The n-BFM is considered very efficient, 
complementary to the MD method, in studying gel electrophoresis. 

In the next section we describe the n-BFM in detail. One further 
detail in algorithm is explained in Appendix \ref{RPBC}. 
The results simulated by the n-BFM are presented and compared with 
those obtained by the c-BFM in Section \ref{sec:results}, and 
the last section is devoted to concluding remarks of the present work.




\section{Method}

In the present work we consider a $d=2$ square lattice version of 
the bond fluctuation method (BFM) \cite{carkre} to simulate gel 
electrophoresis. In this method each monomer is represented by 
a unit cell ($4$ lattice sites) and 
each bond has a variable length $l$ but 
with the restriction $2\le l \le \sqrt{13}$ (see
Fig. \ref{fig:setumei} in Appendix \ref{RPBC}). 
No monomer can come to any site which is already occupied by other 
monomers and by obstacles. 
The latter are substituted for gel fibers and each of them consists
also of 
a unit cell. In the present work they are distributed periodically 
in both $x$ and $y$ directions with period of $a\text{lattice
  units}$.
A uniform, steady electric field is applied in the 
$(1,1)$ direction. We use dimensionless electric field strength
$E=q{\cal E}e/k_BT$, where $q$, 
${\cal E}$, 
$e$, $k_B$ and $T$
denote charge of a monomer, bare electric field strength, lattice 
constant ($\sim$ persistent length, and is put unity 
in the present work),
Boltzmann constant and temperature, respectively. 
The MC processes in the BFM consist of local random walks (of unit 
length) of each monomers in a potential due to the electric field.  
The excluded volume 
effect is satisfied and  no bond crossing 
occurs in this BFM, which we call the conventional BFM (c-BFM). 
Actual simulations are done on square lattices of sizes 
$3M\times 3M$ with periodic boundary conditions. Here $M$ is 
the number of monomers in the chain. 

\begin{figure}[htbp]
  \begin{center}
    \scalebox{0.65}{\includegraphics{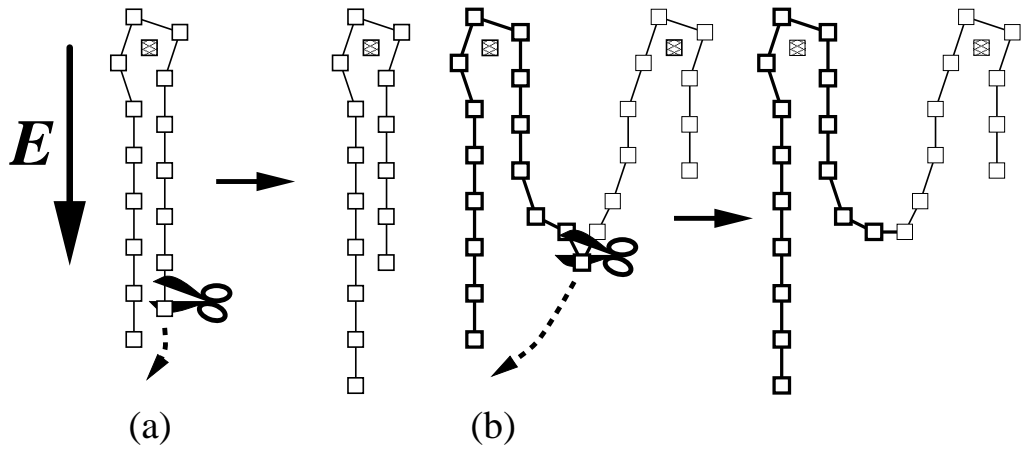}}
    \caption{Schematic illustration of (a) the U-shaped and (b)
      M-shaped conformations pulled down by the field. }
    \label{fig:mosi}
  \end{center}
\end{figure}

Although the c-BFM above described is shown its efficiency in examining 
gel electrophoresis 
in smaller fields, it has serious 
difficulty in larger fields \cite{kremer}. The latter is easily 
understood from the inspection of a case shown in Fig. \ref{fig:mosi}a
where we show a schematic picture 
of the U-shaped conformation of a chain 
hooked by an obstacle and pulled by a large downward field. 
Within the c-BFM 
the chain can escape from the trap only by the following process: 
a relatively slack part of the chain, created around the end segment,
climbs up sequentially against a potential slope of the field up to 
the position of the obstacle. The probability for this process to 
occur is the smaller for the larger field. 
For example, the mobility $\mu$ of the $M=200$ chain simulated by 
the c-BFM increases with increasing $E$ up to 
$E \simeq 0.01$, but it starts to decrease and tends to vanish when 
$E$ is further increased. Here the mobility $\mu$ is simply evaluated 
by
\begin{equation}\label{eqn:mu}
\mu=\frac{\langle X_{\rm G}(t_f)-X_{\rm G}(t_i) \rangle}{t_f-t_i}/E,
\end{equation}
in our units, where $\langle\cdots\rangle$ indicates 
the average over the MC runs. In Eq. (\ref{eqn:mu}) $X_{\rm G}$ is the 
displacement of center of mass in the field direction. 
The starting time $t_i$ of observation is 
chosen to eliminate influence of an initial conformation, and 
$t_f$ is chosen by the condition that 
$\langle X_{\rm G}(t_f)-X_{\rm G}(t_i) \rangle$ is larger than 
$c_\mu a$ or $c_\mu R_{\rm I}$ with $c_\mu \gtrsim 10$, where 
$R_{\rm I}$ is radius of 
gyration of the chain.

In actual experiments DNA can slide off an obstacle even if 
it is temporarily trapped by the latter. 
For example, Volkmuth and Austin fabricated
quasi-two-dimensional microlithographic array of posts and looked at 
a sequence of moving images of 100 kb DNA
in steady field of 
$1.0$v/cm (corresponding to $E\simeq0.005$) \cite{voau}. 
They observed that the DNA hooks one of the 
posts, is extended to nearly its full contour length, and then 
slides from the post. This indicates that tensile force plays an 
important role in such sliding process because the chain is fully 
extended in the process. 

It is then natural to introduce to the c-BFM the following 
non-local processes which simulate realistic sliding motions of 
DNA due to tensile force. In case of the U-shaped conformation 
of Fig. \ref{fig:mosi}a, we simply remove the end monomer of 
the shorter arm and join it to 
the other end of the chain. For the M-shaped conformation shown 
in Fig. \ref{fig:mosi}b, we remove one of the monomers in a dangling 
part at the center to the end of 
the longer arm.

Starting from the above intuitive idea, we have constructed an 
algorithm in which non-local dynamics of monomers is 
systematically and stochastically introduced. 
For this purpose we first define monomers, on which such 
non-local processes are tried. 
They should be in a slack part of the chain, through 
which tensile force cannot transmit. We call them slack monomers 
({\it s}-monomers). 
The {\it s}-monomers defined consist of monomers at both 
ends of a chain and of 
monomers whose nearest neighboring monomers are not separated from 
each others by distance larger than or equal to 4. 
Parts of the chain between neighboring {\it s}-monomers are regarded as tight 
parts, through which tensile force transmits. They are 
considered to slide either partially or as 
a whole depending on non-local movement of {\it s}-monomers which we 
next introduce. 

\begin{figure}[htbp]
  \begin{center}
    \scalebox{0.65}{\includegraphics{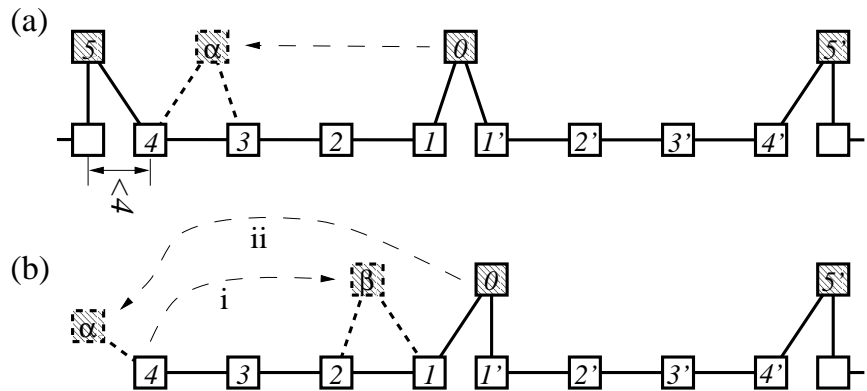}}\vspace*{5mm}
    \caption{Examples of {\it s}-monomers and their movements. (a) The
   case in which  no end monomer is involved.  (b) The case  in
   which an end monomer is involved.
      }
    \label{fig:acc}
  \end{center}
\end{figure}

Our guiding principle to specify a procedure of non-local movement of
{\it s}-monomers is to choose such stochastic processes that any {\it
  s}-monomer moves so as to fulfill detailed balance 
condition in equilibrium. This restriction, however, does not specify a 
procedure uniquely. Among possible procedures we adopt the 
following one:
\begin{enumerate}
  \renewcommand{\labelenumi}{\roman{enumi})}

  \vspace*{-2mm}
  \item Choose one {\it s}-monomer [monomer 0 
in Figs. \ref{fig:acc}a and b shown for examples].

  \vspace*{-2mm}
  \item Count the number of pairs $n$, on which we can try to move 
the chosen {\it s}-monomer, and which are in a region between its 
neighboring {\it s}-monomers including ends of the chain [pairs 12, 23,
 34, and 45 for case 2a 
(`end' for case 2b) between monomers 0 and 5 (`end'), and 1'2', 2'3', 
3'4', and 4'5' between monomers 0 and 5', and so $n=8$]. 

  \vspace*{-2mm}
  \item Choose randomly one of these pairs [$34$ (`end')] with the 
probability $1/n$. 

  \vspace*{-2mm}
  \item Choose randomly one of the allowed conformations putting 
a monomer between the chosen pair [$3\alpha 4$ 
(4$\alpha$)] with the probability $1/W$.  By construction the moved 
monomer is an {\it s}-monomer. 

  \vspace*{-3mm}
  \item Accept the movement of the {\it s}-monomer 
[from {\it s}-monomer 0 to {\it s}-monomer $\alpha$] 
according to the weight $w(\Delta X)$ defined by 
$$ w(\Delta X) = { {\rm e}^{E\Delta X } \over
  {\rm e}^{- E\Delta X } + {\rm e}^{E\Delta X } }, \eqno{(2)} $$
where $\Delta X$ is the displacement of the $s$-monomer in the 
(positive) direction of the field.

\end{enumerate}

\vspace*{-2mm}

\noindent
In (iv) above, the value of $W$ is chosen 
to be greater than max$\{W_1,..,W_n\}$ where $W_i$ is the number of 
the allowed conformations putting a monomer to the $i$-th pair. 
In the present work we fix $W=23$ which is the possible maximum value 
of $W_i$ when the $i$-th pair is one of the `ends'. Then the 
probability of the movement of the $s$-monomer to one specified 
conformation is given by $1/nW$ multiplied weight $w(\Delta X)$ , 
while that of its reversed process is $1/nW$ multiplied 
by weight $w(-\Delta X)$. 
This guarantees the detailed balance condition. 
It is noted here that the acceptance ratio of one non-local movement
without specifying the moved conformation at all, $r_{\rm move}$, is
given by $r_{\rm move} \simeq \overline{W_i}/2W $ in the limit 
$E\Delta X \rightarrow 0 $, where $\overline{W_i}$ is the average of
$W_i$'s including the further restriction below described.
 

There remain, however, two problems which require further 
considerations. One is concerned with the detailed balance 
condition. By the procedure described above it may happen that 
the number of {\it s}-monomers changes depending on positions 
of the monomers around the chosen pair 
[in case Fig. \ref{fig:acc}a, besides the $\alpha$-monomer, monomers 3 
or 4 may become an {\it s}-monomer]. If this happens, the number 
$n$ for the moved {\it s}-monomer also changes after the movement
(except for the case that the movement involves an `end' $s$-monomer).
Then, by the above procedure the probability of the reverse process
violates the detailed balance condition. We get rid of this problem
simply by accepting only processes, by which the number $n$ for the
moved {\it s}-monomer is conserved. 

The other problem remained is how to exclude possible bond crossing 
which may occur when an {\it s}-monomer is moved to an end of the 
chain. We are faced to the same problem when we try to construct 
an initial conformation of a chain by the self-avoiding random walk.  
We have 
developed an algorithm which recognize whether 
bond crossing occurs or not by referring to only local data. It is 
explained in Appendix \ref{RPBC}.

The new BFM (n-BFM) we propose is a combined process of non-local
movements of {\it s}-monomers by the above procedure and the c-BFM 
process\footnote{For a local movement of monomers in the c-BFM in the 
present work, we have adopted the same weight $w(\Delta X)$ with 
$\Delta X = \pm 1/\sqrt2$ after choosing one of the four neighboring 
sites to move with equal probability.} :
each odd (even) MC steps of the c-BFM are followed by trials of 
non-local movements of odd (even) numbered {\it s}-monomers.



\section{Results}\label{sec:results}

\begin{figure}[htbp]
  \begin{center}
    \scalebox{0.65}{\includegraphics{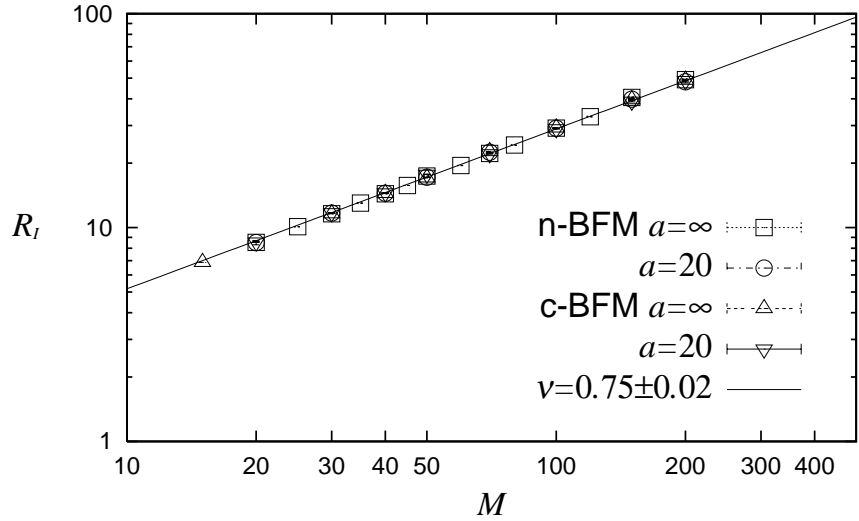}}
    \caption{Dependence of radius of gyration $R_{\rm I}$ on
   length $M$.} 
    \label{fig:T20.RAD}
  \end{center}
\end{figure}

Let us begin with comparison of equilibrium properties in $E=0$ 
simulated by both the c-BFM and the n-BFM. 
As for a representative of static equilibrium properties we show 
four sets of data of radius of gyrations $R_{\rm I}$ in
Fig. \ref{fig:T20.RAD} in cases 
with obstacles ($a=20$) and without obstacles ($a=\infty$) obtained 
by the two BFM's. They all coincide with each others and fit well 
with the power law $R_{\rm I}\propto M^\nu$ with 
$\nu=0.75\pm0.02$ which 
is consistent with the previous results \cite{mcken}. 
There is no significant difference between the cases $a=20$ and
$a=\infty$ in $R_{\rm I}\text{-}M$ dependences. This is because the
concentration of obstacles is too small to change the value of
exponent $\nu$.
\cite{zimmlev}.

\begin{figure}[htbp]
  \begin{center}
    \scalebox{0.65}{\includegraphics{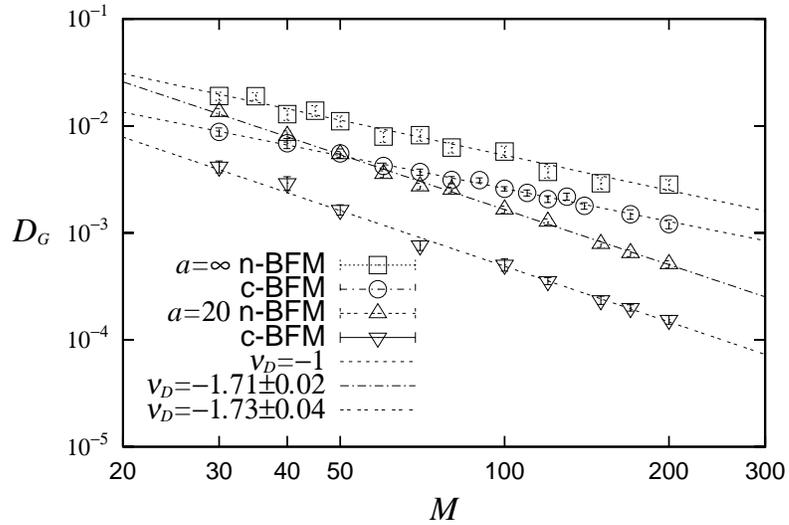}}
    \caption{$D_{\rm G}\text{-}M$ plots for $a=20$, $\infty$ simulated
   by the two BFMs. $\nu_D$ denotes the exponent of the power law
   dependences. }
    \label{fig:DIC}
  \end{center}
\end{figure}

In Fig. \ref{fig:DIC} the diffusion constants $D_{\rm G}$ are plotted
against $M$ to compare behaviors of fluctuation simulated by the two
BFM's. Here $D_{\rm G}$ is evaluated simply by
\begin{equation}
D_{\rm G}=\frac{(\Delta R_{\rm G})^2}{2(t_f-t_i)} \ \ \ \ {\rm with}\ \ \ 
(\Delta R_{\rm G})^2 = \langle \left[ {\bf R}_G(t_f) - {\bf R}_G(t_i)
  \right]^2\rangle, 
\end{equation}
where ${\bf R}_{\rm G}$ is coordinate of the center of mass of the 
chain. The data actually used are those in the range 
$R_{\rm I} \lesssim \Delta R_{\rm G} \le c_{\rm D}R_{\rm I}$ where 
$c_{\rm D}$ is at least larger than 2. 
In case without obstacles both BFM's yield $D_{\rm G} \propto M^{-1}$ 
as expected, while in case $a=20$ the same power law dependence
$D_{\rm G}^{-\nu_{\rm D}}$ with $\nu_{\rm D}=1.71\pm0.02$ (n-BFM) 
and $1.73\pm0.04$ (c-BFM) is obtained. However $D_{\rm G}$ by the 
n-BFM is larger by factor $2\sim3$ than that by the c-BFM. 

\begin{figure}[htbp]
  \begin{center}
    \scalebox{0.65}{\includegraphics{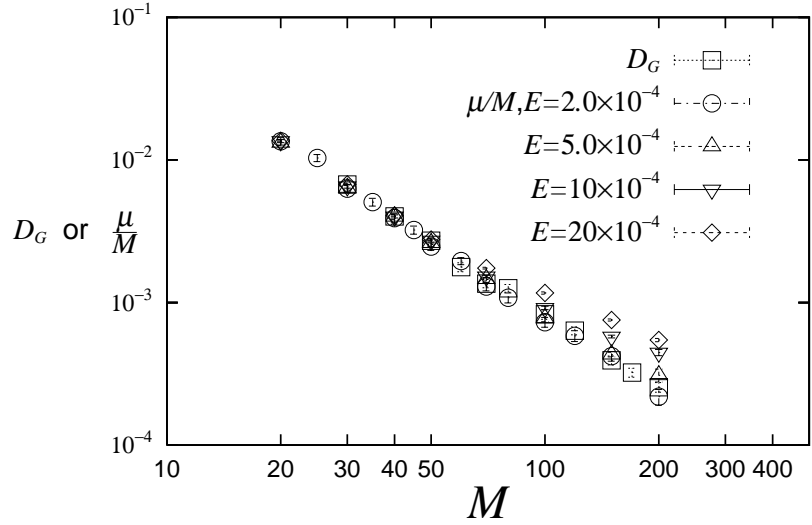}}
    \caption{Diffusion constant $D_G$ and $\mu/M$ plotted against $M$.
   } 
    \label{fig:D_GMU-M}
  \end{center}
\end{figure}

A more stringent check of the new algorithm may be whether it 
reproduces the fluctuation-dissipation theory, or the Einstein 
relation 
$\mu/M = D_{\rm G}$,
where $\mu$ is the mobility in the limit $E\rightarrow0$. 
We plot $\mu/M$ for various $E$ and $D_{\rm G}$ against $M$ in
Fig. \ref{fig:D_GMU-M}. Clearly we see the Einstein relation holds well. 
The results described so far indicate that our new algorithm 
reproduces equilibrium properties of a polymer quite satisfactorily.

\begin{figure}[htbp]
  \begin{center}
    \scalebox{0.65}{\includegraphics{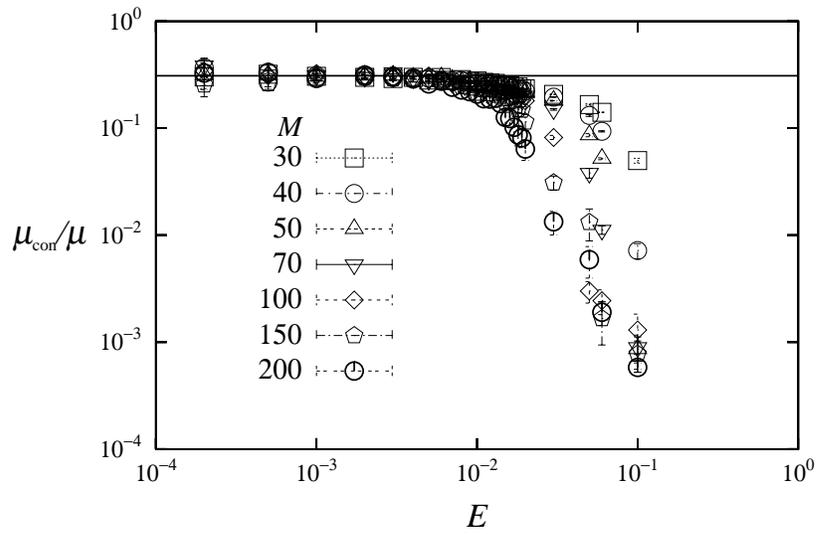}}
    \caption{The mobility ratio $\mu_{\text{con}}/\mu$ plotted against
      $E$. Where $\mu_{\text{con}}$ is obtained by the c-BFM and 
      $\mu$ by the n-BFM.
}
    \label{fig:tscale}
  \end{center}
\end{figure}

Let us next examine stationary properties in finite $E$ simulated. 
In order to compare those obtained by the two BFM's, we show the 
ratio $\mu_{\rm con}/\mu$ for various $M$  with $a=20$
plotted against $E$ in Fig. \ref{fig:tscale}. Here $\mu_{\rm con}$ and 
$\mu$ are mobilities simulated by the c-BFM and the n-BFM, 
respectively. In small fields $E \lesssim 5.0\times 10^{-3}$ the ratio 
is almost constant and is also  independent of $M$. The constant 
value is equal to the ratio of the diffusion constants $D_{\rm G}$ 
of the two BFM's mentioned above. 

\begin{figure}[htbp]
  \begin{center}
    \scalebox{0.65}{\includegraphics{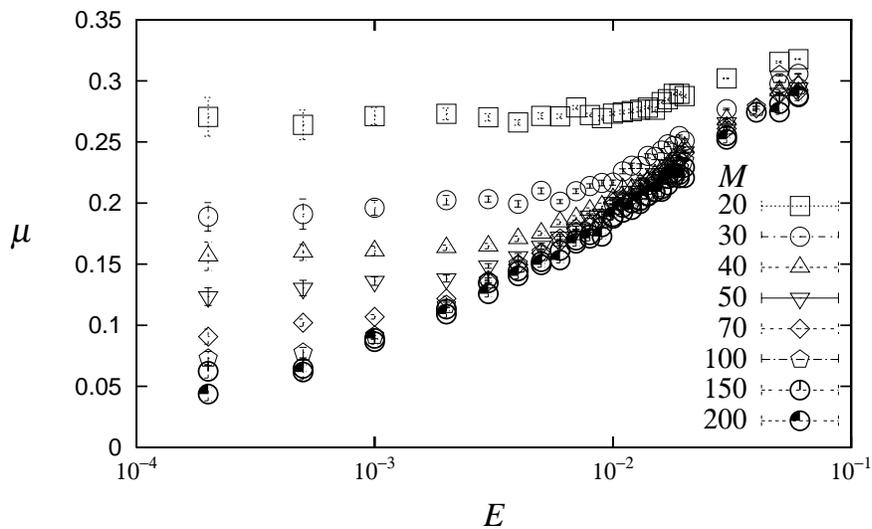}}
    \caption{Plots of $\mu$ vs. $\log E$. }
    \label{fig:MU-E}
  \end{center}
\end{figure}

The ratio $\mu_{\rm con}/\mu$ deviates from the constant 
value when $E$ exceeds a certain value $E_{\rm th}$ which depends on 
$M$. Its deviation is more significant for chains with larger $M$, 
and the ratio tends to vanish at $E$ much larger than $E_{\rm th}$. 
This result simply reflects the difficulty 
of the c-BFM described in the previous section and is considered 
to have nothing to do with the n-BFM. Actually there seems no 
crossover-like behavior around $E_{\rm th}$ in $\mu$ obtained by the 
n-BFM as shown in Fig. \ref{fig:MU-E}. One sees in the figure that
$\mu$ of each $M$ increases monotonically with increasing $E$, and
tends to saturate at largest $E\ (\sim 0.1)$ we have simulated. 
Thus the data in Figs. \ref{fig:tscale} and \ref{fig:MU-E} tell us that 
we have in fact overcome the difficulty of the c-BFM by means of 
the n-BFM we have proposed. 

\begin{figure}[htbp]
  \begin{center}
    \scalebox{0.65}{\includegraphics{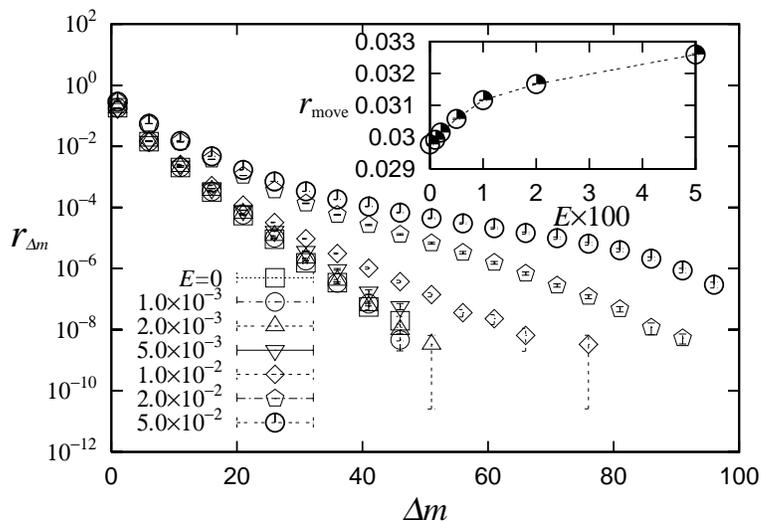}}
    \caption{ Histograms of
   the $s$-monomer movements accepted, $r_{\Delta m}$, against the 
   distance (along the chain) $\Delta m$ for the chain with $M=100$ and 
   $a=20$. In the inset is shown the acceptance ratio of the 
   $s$-monomer movement, $r_{\rm move}$. } 
    \label{fig:accept}
  \end{center}
\end{figure}

In order to get further insights of the n-BFM which gives rise to 
quite satisfactory results so far demonstrated, we have examined some 
details of the $s$-monomer movement in a system with $M=100$ and 
$a=20$. Interestingly, the acceptance ratio of the $s$-monomer 
movement, $r_{\rm move}$, defined in the previous section is 
rather small ($\sim 3\%$) and does not depend sensitively on $E$ 
in the range examined as shown in the inset of Fig. \ref{fig:accept}. 
On the other hand, the average number of the 
$s$-monomers, ${\bar M_{\rm s}}$, is rather large: 
${\bar M_{\rm s}} \simeq 28$ at $E=0$ and ${\bar M_{\rm s}}\simeq 24$
at $E=0.05$. 

These results are interpreted as follows. In the field range 
investigated here ($E \le 0.05$), the entropy effect dominates 
to determine conformations of a chain. The latter is therefore in a 
coiled conformation on average. The $s$-monomer movement in this 
situation is regarded as local fluctuation of the conformation due to 
the entropy effect rather than due to the tensile force. Although 
$r_{\rm move}$ is rather small, the $s$-monomer movements 
contribute to displacement of the center of mass of the chain which is 
comparable in magnitude with that by one mcs of the c-BFM process. 
This explains why $D_{\rm G}$ and 
$\mu$ by the n-BFM are larger than the corresponding 
quantities by the c-BFM (note that one mcs in the n-BFM consists 
of one mcs of the c-BFM process and trials of non-local movement 
on half of the $s$-monomers). Such efficiency of the $s$-monomer 
movement to $D_{\rm G}$ and $\mu$ may be attributed to the fact 
that the $s$-monomer moves along the chain 
direction which is hardly realized by the c-BFM process. 


The main part of Fig. \ref{fig:accept} is the histogram of the
relative ratio of 
the $s$-monomer movements accepted, $r_{\Delta m}$, against the 
distance (along the chain), $\Delta m$, they moved. Notice that the 
movement with $\Delta m \sim M/2$ occurs even at $E=0$, though 
$r_{\Delta m}$ is quite small (the abscissa is in the logarithmic 
scale). This $r_{\Delta m}$ at $E=0$ reflects how frequently, 
in fact quite rarely,  
extended conformations occur by fluctuation in the coiled
conformation of the chain. 

An important observation of Fig. \ref{fig:accept} is that $r_{\Delta
  m}$ for 
$E\ge 0.02$ is much enhanced at large $\Delta m$, though the absolute 
magnitude is still quite small. This means that even in this range of 
$E$ extended conformations occur with low probability. But once 
they are present, the $s$-monomer movements with large $\Delta m$ 
occur with relatively high probability. This situation is just 
what we have first intuitively expected: the tensile force is 
considered to play an important role on dynamics of the chain. The low
probability 
of such non-local movements of the $s$-monomer, which nevertheless
overcomes the difficulty in the c-BFM and reproduces smooth, 
monotonic $E$-dependence of the mobility, justifies our introduction 
of the $s$-monomer movements as a stochastic process.

\begin{figure}[htbp]
  \begin{center}
    \scalebox{0.65}{\includegraphics{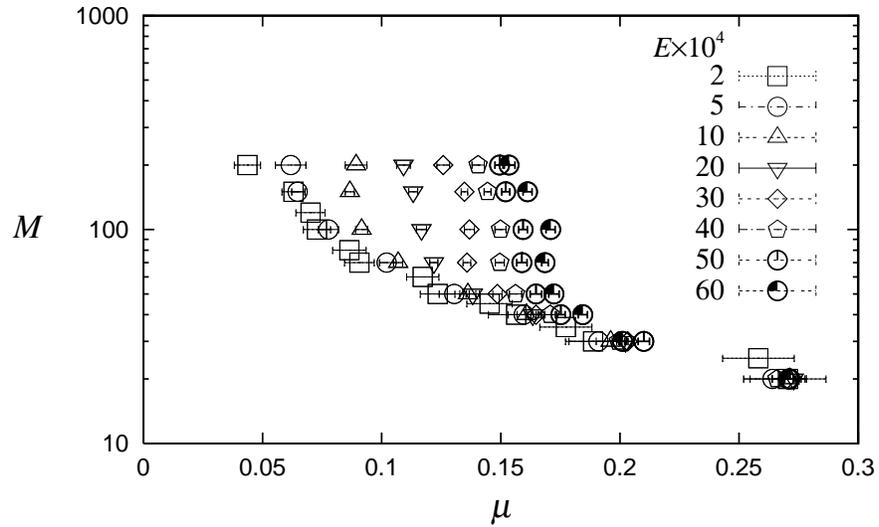}}
    \caption{Plots of $\log M$ against $\mu$. }
    \label{fig:MU-M}
  \end{center}
\end{figure}

In Fig. \ref{fig:MU-M} we show the $\mu$ versus $\log M$ plot using the
same data shown in Fig. \ref{fig:MU-E}.  
It can be seen that $\mu$ for each $E$ decreases monotonically as $M$
increases, and that they are strongly dependent on $M$ when $E$ is 
small enough. As $E$ increases, however, it gradually looses 
$M$-dependence from larger values of $M$, and finally $\mu$ becomes 
almost independent of $M$ when $E$ is quite strong ($E\sim 0.1$). 
These behavior of $\mu$ is qualitatively consistent with 
experimental results \cite{hebe}.

\begin{figure}[htbp]
  \begin{center}
    \scalebox{0.65}{\includegraphics{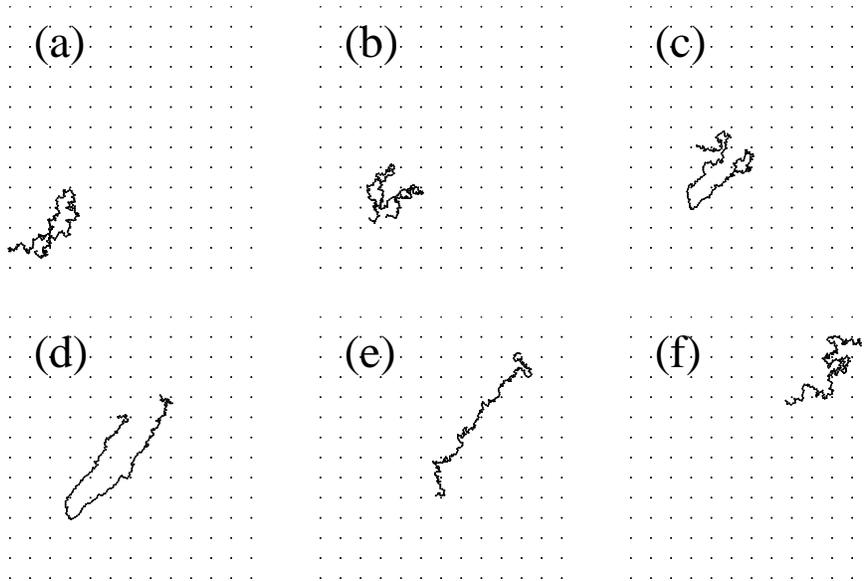}}
    \caption{Time development, (a), (b), \dots, (f), of $M=200$ chain
   conformations in field ${\bf E}\parallel(1,1)$ and $E=1.02\times 10^{-2}$. 
      The snapshots at every $4.0\times 10^{4}\text{mcs}$ are shown. Grid
      points represent obstacles ($a=30$). }
    \label{fig:inch-worm}
  \end{center}
\end{figure}

Finally we note that chain conformation dynamics simulated by the 
present n-BFM is actually complicated. Particularly in a certain 
limited range of $E \ (\sim 0.01)$ sequences of contraction and 
extension are observed. One of such sequences are shown in
Fig. \ref{fig:inch-worm} which are snapshots of conformation of an
$M=200$ chain in every $4\times10^{4}$ mcs in 
a certain MC run. A chain moving in a relatively collapsed 
form (a) is trapped by a certain obstacle (b), deforms to a V- or 
U-shaped conformation (c,d), slides off the obstacle (e), and then 
tends to form a collapsed conformation again (f). 
In contrast to experimental observation \cite{oana}, however, 
quasi-periodical alternation between contracted and extended 
conformations has not yet been observed, or periods between the two 
conformations are rather random. 



\section{Concluding Remarks}

By introducing stochastic, non-local movements of slack parts of the
polymer ($s$-monomers) into the conventional bond fluctuation method
(c-BFM), we have constructed a new BFM (n-BFM) algorithm which
overcomes the 
difficulty of the c-BFM applied to gel electrophoresis in relatively
large field, namely, polymers once hooked by gel fibers become unable
to get rid of them. In smaller fields, on the other hand, the new
stochastic process gives rise to conformational change of the polymer
which is interpreted as the ordinary entropy effect. 
The present n-BFM thus turns out to be able to reproduce qualitative 
aspects of gel electrophoresis phenomena in a wide range of the field.

The n-BFM is considered more effective for a denser gel (a smaller
$a$). Actually the preliminary analysis of the mobility in case $a=12$ 
yields $\mu_{\rm con}/\mu \simeq 0.25$ at small $E$, which is smaller
than that of the case $a=20$ shown in Fig. \ref{fig:tscale}. This
tells that non-local $s$-monomer movements make it easier for parts of
polymer to escape obstacles, and so even the entropic conformation
change in smaller fields is fastened as compared with the c-BFM.

There are many problems to be explored further. Among them are, an
improvement of algorithms to increase the acceptance ratio of the
$s$-monomer movements, $r_{\rm move}$, without violating the detailed
balance, and quantitative comparisons with experimental observations. 
For the latter purpose we have to adopt model systems
with more realistic distribution and/or shape of obstacles, and to
extend the algorithm to $d=3$ system. These problems are now under
investigations.

\section*{Acknowledgments}

We thank Y. Masubuchi and M. Doi for stimulating discussions 
on their experimental and theoretical works on gel electorophoresis. 
We also acknowledge useful discussions with S. Todo and K. Hukushima.
The computation in this work has been done using the facilities of 
the Supercomputer Center, Institute for Solid State Physics, 
University of Tokyo, and those of the Computer Center of University 
of Tokyo.
\appendix

\section{Rejection procedure for bond crossing}\label{RPBC}

\begin{figure}[htbp]
  \begin{center}
    \scalebox{0.65}{\includegraphics{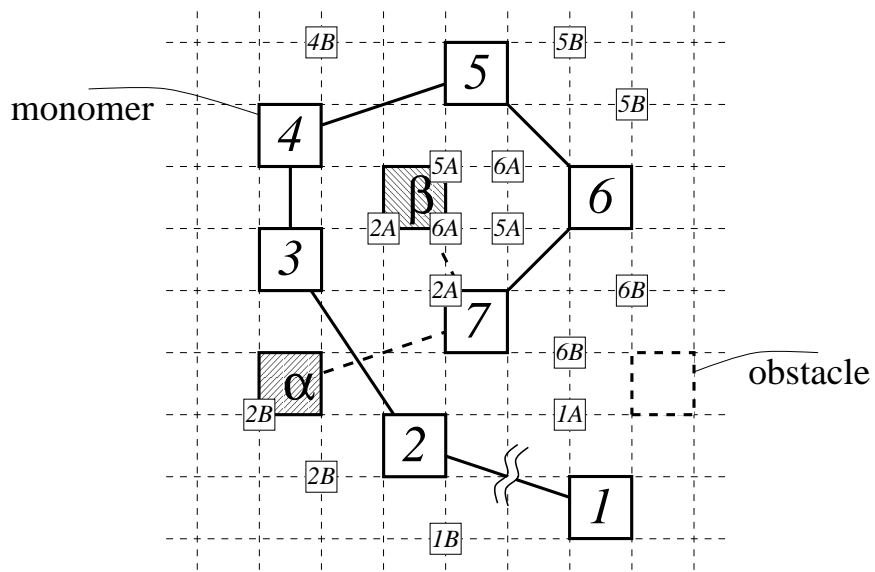}}
    \caption{An example of conformations of a $M=7$ BFM chain 
   on a square lattice. The unit square with thick broken lines
   represents an obstacle. Small 
      squares on sites denote the local function $\Phi(n)$.
      }   
    \label{fig:setumei}
  \end{center}
\end{figure}

Here we explain a method how to reject a non-local movement of 
an {\it s}-monomer to one end of a chain which associates with 
bond crossing. For this purpose we use, as a simple example, 
an $M=7$ BFM chain drawn in Fig. \ref{fig:setumei}. In addition to monomers and 
connecting bonds, we introduce a local 
function $\Phi (n)$ 
represented by small squares on each site $n$. This function
contains information on position of each site $n$ relative to 
nearby bonds which are now specified as vectors 
$\overrightarrow{i,i+1}$. When $\Phi (n)$ takes a value `$i$A' 
(`$i$B'), it means that site $n$ is near 
the bond $\overrightarrow{i,i+1}$ (both distances from monomers $i$ 
and $i+1$ are less than 4) and is in its wright (left) hand side. 
Since site $n$ can be near other bonds $\Phi (n)$ is, in general, 
multi-valued. If, on the other hand, site $n$ is far from any bonds 
$\Phi (n) = $`null'.

By making use of the local function $\Phi (n)$ thus defined we can 
check bond crossing as follows. To be simple, let us consider the 
case removing end monomer 1 and putting it to the other end monomer 
$7$, as shown in Fig. \ref{fig:setumei}. 
If monomer 7 touches a site with $\Phi = $`$2A$', 
the moved monomer should not touch sites with $\Phi = $`$2B$'. 
Monomer $\alpha$ is such a case where bonds $\overrightarrow{2,3}$
and $\overrightarrow{7,\alpha}$ cross and so is rejected. 
Other cases such as monomer $\beta$ are all accepted from the 
present criterion on bond crossing.






\end{document}